\documentclass[preprint,12pt]{elsarticle}

\usepackage{amssymb,mathrsfs,booktabs}
 \usepackage{amsthm}
 \usepackage{xcolor}
 \usepackage{booktabs}
\usepackage[flushleft]{threeparttable}
\usepackage[font=small,labelfont=bf]{caption}
\newcommand{\tabincell}[2]{\begin{tabular}{@{}#1@{}}#2\end{tabular}}

\newtheorem{theorem}{Theorem}[section]
\newtheorem{lemma}{Lemma}[section]
\newtheorem{corollary}{Corollary}[section]

\newtheorem{remark}{Remark}[section]
\newtheorem{definition}{Definition}[section]
\newtheorem{proposition}{Proposition}[section]
\newtheorem{example}{Example}[section]
\newtheorem{assumption}{Assumption}[section]

\journal{Journal}

\begin{document}

\begin{frontmatter}



\title{Reachability of Dimension-Bounded Linear Systems\tnoteref{thanks}}

\tnotetext[thanks]{This work was supported by the National Natural Science Foundation of China (11701258, 61379021, 61773371, 61877036, 11871259), and the Natural Science Foundation of Shandong Province (ZR2019MF002).}


\author[a]{Yiliang Li}
\ead{liyiliang1994@126.com}
\author[b]{Haitao Li}
\ead{haitaoli09@gmail.com}
\author[a]{Jun-e Feng\corref{corr1}}
\ead{fengjune@sdu.edu.cn}
\author[c]{Jinjin Li}
\ead{jinjinli@mnnu.edu.cn}

\cortext[corr1]{Corresponding author: Jun-e Feng}

\address[a]{School of Mathematics, Shandong University, Jinan, Shandong 250100, P.R. China}
\address[b]{School of Mathematics and Statistics, Shandong Normal University, Jinan, 250014, P.R. China}
\address[c]{School of Mathematics and Statistics, Minnan Normal University, Zhangzhou, Fujian 363000, P.R. China}


\begin{abstract}
In this paper, the reachability of dimension-bounded linear systems is investigated.
Since state dimensions of dimension-bounded linear systems vary with time, the expression of state dimension at each time is provided.
A method for judging the reachability of a given vector space $\mathcal{V}_{r}$ is proposed.
In addition, this paper proves that the $t$-step reachable subset is a linear space, and gives a computing method.
The $t$-step reachability of a given state is verified via a rank condition.
Furthermore, annihilator polynomials are discussed and used to illustrate the relationship between the invariant space and the reachable subset after the invariant time point $t^{\ast}$.
The inclusion relation between reachable subsets at times $t^{\ast}+i$ and $t^{\ast}+j$ is shown via an example.
\end{abstract}


\end{frontmatter}


\section{Introduction}
Cross-dimensional systems are also called dimension-varying systems or dimension free systems \cite{Chengdaizhan2019b}.
Some mathematic models with different dimensions can be described as cross-dimensional systems, such as biological systems \cite{Huangronghua1995}, electric power generators \cite{MachowskiJ1997} and vehicle clutch systems \cite{Chengdaizhan2020a}.
Four phenomena appeared in the spacecraft formation \cite{Panjiao2014}, docking, undocking, departure and participation, are also practical examples of cross-dimensional systems.
Take participation as an example.
Some new spacecrafts join in the formation, or the departed spacecrafts come back.
In these cases, their states are treated as new ones and considered in the next mode.
Thus, state dimensions increase in the next mode.
For simulating the whole flying process of spacecraft formation, a switched system approach was used to handle dimension-varying systems \cite{Yanghao2013}.
In the spacecraft formation, it is important to guarantee a smooth transition from one mode to another, which requires that the system is continuous at the mode conversion time.
But the continuity of switched systems at the switching time cannot be determined.
Hence, it is necessary to find more ideal models for modeling spacecraft formation.

Hybrid systems, reflected the interaction of continuous- and discrete-time dynamics \cite{GoebalRafal2009}, can be also applied to study dimension-varying systems.
From the evolution of hybrid systems, there are at least two subsystems in a hybrid system.
When an event occurs, the hybrid system switches from the current discrete mode to a new discrete mode.
The switching rule is provided via a given reset map \cite{Habets2006}.
Consider a discrete dynamic model generating a sequence of modes as a switching signal defined in a switched system.
Hybrid systems can be viewed as switched systems \cite{LiberzonD2014}.
Thus, the most basic way to deal with dimension-varying systems is to switch.
Namely, the continuity at the dimension change time is also not determined via the hybrid system method.
To solve this problem, a unified form model should be established for dimension-varying systems.
Motivated by it, cross-dimensional linear systems were presented by Prof. Cheng \cite{Chengdaizhan2018d}.

The difficulty of giving a unified form model for dimension-varying systems is how to connect spaces with different dimensions together.
Thanks to Cheng operations, this difficulty was solved, and cross-dimensional linear systems were established.
The Cheng operations include semi-tensor product of matrices, M-addition of matrices, V-addition of vectors and V-product of matrices and vectors \cite{Chengdaizhan2019c}.
Using these operations, cross-dimensional linear systems can go cross spaces with different dimensions \cite{Chengdaizhan2018d}.
The next problem is how to apply cross-dimension systems to handle the dynamics of the transient process of practical examples, most of which have invariant dimensions except the transient period.
In the light of the proposed projection among spaces with different dimensions \cite{Chengdaizhan2018e}, \cite{Chengdaizhan2020a} presented a technique to realize the dimension transient process, and provided an example to show this design technique.
From the given example, one sees how to determine the continuity at the dimension change time.

The reachability analysis of a dynamical system refers to compute a reachable set, which contains the entire state trajectories of the system starting from uncertain initial conditions and driven by uncertain inputs.
Up to now, there are many references concerning the reachability analysis of dynamical systems, such as
linear systems \cite{Dubaozhu2016,Alfredorrnarvaez2016,Hugonestorvillegaspico2018},
switched systems \cite{Simonebaldi2018,Chenyong2016,Feizhongyuan2018},
hybrid systems \cite{Granatog2014,Moussamaiga2016}, logical control networks \cite{Liyalu2019a,Liuyang2020a,Zouyunlei2016a} and other systems \cite{Fengzhiguang2015,FornasiniE2011,Magronvictor2019}.
Notably, \cite{Bokanowskiolivier2016,Guoyuqian2018a} analyzed the observability and optimal control problems via the reachability approach.
Besides, the reachability plays an important role in practical problems \cite{Matthiasalthoff2014,Hugonestorvillegaspico2016,Sujinya2019}.
Since the reachability of dynamical systems occupies a significant position in both theoretical developments and practical applications,
it is meaningful to study the reachability of cross-dimensional linear systems.

Because state dimensions of cross-dimensional linear systems vary with time,
the first step of taking the reachability into account is to discuss state dimensions of the system.
\cite{Chengdaizhan2018d} pointed out that cross-dimensional linear systems are classified into two cases: dimension-unbounded linear systems and dimension-bounded linear ones.
For the former one, after a certain time, state dimensions not only increase with time but also go to infinity \cite{Chengdaizhan2018d}.
Thus, the increase time and state dimensions of dimension-unbounded linear systems were studied in \cite{Fengjune2021a}.
For the later one, state dimensions are invariant after a certain time $t^{\ast}$, which is called the invariant time point.
It means that the trajectory of the system enters to an invariant space.
A recursive formula presented by \cite{Chengdaizhan2018d} was used to compute the dimension of state at each time.
However, using this recursive formula, the dimension of state at time $t$ is obtained, when dimensions of states before time $t$ are all calculated.
Hence, giving an approach to computing state dimensions directly draws our attention.
The expression of state dimension after time $t^{\ast}$ was provided by \cite{Zhaopeixin2021}.
But the invariant time point $t^{\ast}$ and state dimensions before time $t^{\ast}$ are not researched.
Furthermore, to the best of our knowledge, there are no results on reachable subsets of dimension-bounded linear systems.

Based on the analysis above, this paper investigates the reachability of dimension-bounded linear systems.
The main contributions of this paper are as follows.

\begin{itemize}
  \item For a given dimension-bounded linear system, the expression of state dimension at each time is provided.
      Compared with the recursive formula proposed by \cite{Chengdaizhan2018d}, it is easier to judge whether a given integer is a reachable dimension according to this result.
      It also reveals the dimension variation law of dimension-bounded linear systems clearly.
  \item By proving that the $t$-step reachable subset is a linear space, this paper concludes that the $t$-step reachable subset equals the span of a list of specific vectors.
      A rank condition is presented to verify the $t$-step reachability of a given state.
      Note that these results also hold for dimension-unbounded linear systems.
  \item For illustrating the relationship between the invariant space and the reachable subset after the invariant time point $t^{\ast}$, annihilator polynomials are discussed.
      The obtained results show that $A$-annihilator of vector space $\mathcal{V}_{n}$ is the generalization of conventional annihilator polynomial, where $A$ is a given matrix.
  \item An example is studied to explain the inclusion relation between reachable subsets at times $t^{\ast}+i$ and $t^{\ast}+j$.
      This example makes the discussion of reachable subset more perfect.
\end{itemize}

The rest of this paper is organised as follows.
Preliminaries and problem formulation are provided in Sections 2 and 3, respectively.
Section 4 is main results of this paper, including the discussion of state dimension, some results about annihilator polynomials and the analysis of reachable subset.
Section 5 gives some concluding remarks.

\section{Preliminaries}

In this section, we provide a list of notations, some results about V-product and V-addition.

$\bullet\ \ \mathcal{M}_{m\times n}$ is the set of all $m\times n$ real matrices.
Denote $\mathcal{M}:=\bigcup\limits_{m=1}^{\infty}\bigcup\limits_{n=1}^{\infty}\mathcal{M}_{m\times n}$.

$\bullet\ \ \mathcal{V}_{n}$ is the set of all $n$ real column vectors.
Denote $\mathcal{V}:=\bigcup\limits_{n=1}^{\infty}\mathcal{V}_{n}$.

$\bullet\ \ \mathbb{N}$ is the set of all non-negative integers.

$\bullet\ \ \mathbb{R}$ is the set of all real numbers.

$\bullet\ \ \mathrm{lcm}(m,n)$ represents the least common multiple of $m$ and $n$.

$\bullet\ \ a\mid b$ means that integer $a$ is a divisor of integer $b$.

$\bullet\ \ \mathbf{0}$ is a null matrix.

$\bullet\ \ \delta_{n}^{i}$ is the $i$th column of identity matrix $I_{n}$.

$\bullet\ \ \mathbf{1}_{m\times n}=[\mathbf{1}_{m}\ \cdots\ \mathbf{1}_{m}]\in\mathcal{M}_{m\times n}$, where $\mathbf{1}_{m}:=\sum\limits_{i=1}^{m}\delta_{m}^{i}$.

$\bullet\ \ $
$\mathrm{span}\{\alpha_{1},\alpha_{2},\ldots,\alpha_{s}\}=\{k_{1}\alpha_{1}+k_{2}\alpha_{2}+\cdots+k_{s}\alpha_{s}|
\alpha_{i}\in\mathcal{V}_{n},\forall k_{i}\in\mathbb{R},i=1,2,\ldots,$
$s\}.$

$\bullet\ \ $For two given matrices $P\in\mathcal{M}_{m\times n}$ and $Q\in\mathcal{M}_{p\times q}$,
the semi-tensor product of $P$ and $Q$ is defined as
\[P\ltimes Q:=(P\otimes I_{t/n})(Q\otimes I_{t/p}),\]
where $t=\mathrm{lcm}(n,p)$, $\otimes$ is the Kronecker product.

As main tools for addressing cross-dimensional linear systems, V-product and V-addition are defined.

\begin{definition}\label{Def1}\rm\cite{Chengdaizhan2019b}
(1) Let $A\in\mathcal{M}_{m\times n}, x\in\mathcal{V}_{r}$ and $s=\mathrm{lcm}(n,r)$.
The V-product of $A$ and $x$, denoted by $\vec{\ltimes}$, is defined as
\[A\vec{\ltimes}x:=(A\otimes I_{s/n})(x\otimes\mathbf{1}_{s/r}).\]

(2) Let $x\in\mathcal{V}_{n},y\in\mathcal{V}_{r}$ and $s=\mathrm{lcm}(n,r)$.
The V-addition 
of $x$ and $y$, denoted by $\vec{\rotatebox[origin=c]{90}{$\mp$}}$, is defined as
\[x\vec{\rotatebox[origin=c]{90}{$\mp$}}y:=(x\otimes\mathbf{1}_{s/n})+(y\otimes\mathbf{1}_{s/r}).\]
\end{definition}

If $n=r$, then $A\vec{\ltimes}x=Ax$ and $x\vec{\rotatebox[origin=c]{90}{$\mp$}}y=x+y$.
That is, V-addition and V-product are generalizations of conventional vector addition and conventional vector product, respectively.
The following lemmas were proved in \cite{Chengdaizhan2019b}.

\begin{lemma}\label{Le2}\rm\cite{Chengdaizhan2019b}
Consider V-product $\vec{\ltimes}:\mathcal{M}\times\mathcal{V}\rightarrow\mathcal{V}$.
It is linear with respect to the second variable, precisely,
$$A\vec{\ltimes}(ax\vec{\rotatebox[origin=c]{90}{$\mp$}}by)=aA\vec{\ltimes}x\vec{\rotatebox[origin=c]{90}{$\mp$}}bA\vec{\ltimes}y, a,b\in\mathbb{R}.$$
\end{lemma}

\begin{lemma}\label{Le3}\rm\cite{Chengdaizhan2019b}
For any two matrices $A,B\in\mathcal{M}$ and any vector $x\in\mathcal{V}$, it holds that
$$(A\ltimes B)\vec{\ltimes}x=A\vec{\ltimes}(B\vec{\ltimes}x),$$
Moreover, we have
$$A^{i}\vec{\ltimes}x=(A\ltimes\cdots\ltimes A)\vec{\ltimes}x=A\vec{\ltimes}(A\vec{\ltimes}\cdots\vec{\ltimes}(A\vec{\ltimes}x)).$$
\end{lemma}

The definition of invariant space is given in the following.

\begin{definition}\label{Def2}\rm\cite{Chengdaizhan2019b}
For a given mattix $A\in\mathcal{M}_{m\times n}$, vector space $\mathcal{V}_{r}$ is called an $A$-invariant space if
$A\vec{\ltimes}x\in\mathcal{V}_{r}$ for any $x\in\mathcal{V}_{r}.$
\end{definition}

\section{Problem formulation}

Consider a cross-dimensional linear system
\begin{equation}\label{Eq3}
x(t+1)=A\vec{\ltimes}x(t), x(0)=x_{0},
\end{equation}
where $A\in\mathcal{M}_{m\times n}$.
Cross-dimensional linear system (\ref{Eq3}) is called a time invariant discrete linear pseudo dynamical system \cite{Chengdaizhan2018d}.
Combined with Definition \ref{Def2}, dimension-bounded linear systems are defined.

\begin{definition}\label{Def3}\rm\cite{Chengdaizhan2018d}
Consider system (\ref{Eq3}). $A$ is called a dimension-bounded operator, if for any $x(0)=x_{0}\in\mathcal{V}_{p}$, there exists a $t^{\ast}>0$ and an $r^{\ast}$ such that $x(t)\in\mathcal{V}_{r^{\ast}}$ holds for any $t\geq t^{\ast}$,
where $\mathcal{V}_{r^{\ast}}$ is called the invariant space of system (\ref{Eq3}).
Additionally, system (\ref{Eq3}) is called a dimension-bounded linear system.
\end{definition}

In this paper, system (\ref{Eq3}) is a dimension-bounded linear system unless otherwise specified.
According to Lemma \ref{Le1}, it is reasonable to assume $A\in\mathcal{M}_{m\times km}$.

\begin{lemma}\label{Le1}\rm\cite{Chengdaizhan2019b}
$A\in\mathcal{M}_{m\times n}$ is dimension-bounded if and only if $m\mid n$.
\end{lemma}

For analyzing the reachability of dimension-bounded linear systems, the following definition is introduced.

\begin{definition}\label{Def4}\rm
Consider dimension-bounded linear system (\ref{Eq3}) with initial space $\mathcal{V}_{p}$.

(1) A state $x$ is said to be $t$-step reachable, if there exists an initial value $x(0)\in\mathcal{V}_{p}$ such that the trajectory of the system reaches $x$ from $x(0)$ at time $t$.

(2) $x$ is said to be reachable, if there exists an integer $t$ such that $x$ is $t$-step reachable.
Moreover, the dimension of $x$ is called a reachable dimension of system (\ref{Eq3}).

(3) $R\subset\mathcal{V}$ is called a reachable subset, if each state $x\in R$ is reachable.
\end{definition}

For a given dimension-bounded linear system with initial space $\mathcal{V}_{p}$, dimensions of $x(t_{1})$ and $x(t_{2})$ may be different.
Thus, the first step of judging whether state $x\in\mathcal{V}_{r}$ is $t$-step reachable is to show the reachability from $\mathcal{V}_{p}$ to $\mathcal{V}_{r}$.
Here the reachability from $\mathcal{V}_{p}$ to $\mathcal{V}_{r}$ means that $r$ is a reachable dimension of the corresponding system.
The second step is to verify whether each state $x\in\mathcal{V}_{r}$ is reachable.
Therefore, state dimensions and the $t$-step reachable subset are discussed in this paper.
Moreover, this paper focuses on the reachability after the invariant time point $t^{\ast}$.
To show the relationship between the invariant space and the reachable subset after time $t^{\ast}$, some results about annihilator polynomials are presented in this paper.
A numerical example is provided for illustrating the inclusion relation between reachable subsets at times $t^{\ast}+i$ and $t^{\ast}+j$.

\section{Main results}

This section studies the reachability of dimension-bounded linear systems from four aspects: state dimension at each time, the $t$-step reachable subset, the relationship between the invariant space and the reachable subset after the invariant time point $t^{\ast}$ and the inclusion relation between reachable subsets at times $t^{\ast}+i$ and $t^{\ast}+j$.
It is worth noting that annihilator polynomials are useful in discussing the relationship between the invariant space and the reachable subset after the invariant time point $t^{\ast}$.
Thus, this section also focuses on annihilator polynomials.

\subsection{The discussion of state dimension}

This subsection proposes a method for judging whether a  given integer is a reachable dimension.
To this end, an example is provided to show the relationship between reachable dimensions and dimensions of system matrix $A$ and initial space $\mathcal{V}_{p}$.

\begin{example}\label{Ex1}\rm
Consider a dimension-bounded linear system $x(t+1)=A\vec{\ltimes}x(t)$
with $A\in\mathcal{M}_{m\times km}$ and initial value $x(0)\in\mathcal{V}_{p}$.
Denote the dimension of state $x(t)$ by $r(t)$.
The change of $r(t)$ is shown in Table \ref{Tab1}.
\begin{table}[!htbp]
\tiny
\caption{\label{Tab1}}
\centerline{
\begin{tabular}{cccc}
\toprule
state dimension & \tabincell{c}{$m=10$\\ $k=6$\\ $p=68040$} & \tabincell{c}{$m=6$\\ $k=20$\\ $p=30$} &\tabincell{c}{$m=48$\\ $k=1715$\\ $p=18900$} \\
\midrule
$r(0)$&$2^{3}\times 3^{5}\times 5\times 7$&$2\times 3\times 5$&$2^{2}\times 3^{3}\times 5^{2}\times 7$\\
$r(1)$&$2^{2}\times 3^{4}\times 5\times 7$&$2\times 3$&$2^{4}\times 3^{3}\times 5$\\
$r(2)$&$2\times 3^{3}\times 5\times 7$&$2\times 3$&$2^{4}\times 3^{3}$\\
$r(3)$&$2\times 3^{2}\times 5\times 7$&$2\times 3$&$2^{4}\times 3^{3}$\\
$r(4)$&$2\times 3\times 5\times 7$&$2\times 3$&$2^{4}\times 3^{3}$\\
$r(5)$&$2\times 5\times 7$&$2\times 3$&$2^{4}\times 3^{3}$\\
$r(6)$&$2\times 5\times 7$&$2\times 3$&$2^{4}\times 3^{3}$\\
$r(7)$&$2\times 5\times 7$&$2\times 3$&$2^{4}\times 3^{3}$\\
$\cdots$&$\cdots$&$\cdots$&$\cdots$\\
\bottomrule
\end{tabular}}
\end{table}

From Table \ref{Tab1}, the following conclusions are derived.

(1) State dimensions before the invariant time point $t^{\ast}$ decrease with time.

(2) One obtains $m\mid r(t)$.

(3) There may be a factor $m_{1}$ of $m$ such that $mm_{1} \mid r(t)$ holds.

(4) Compared $r(t)$ with $r(t-1), t\leq t^{\ast}$, it is easy to see that $r(t-1)=k_{1}lr(t)$, where $k_{1}\mid k, l\in\mathbb{N},l\neq 0$.

(5) If $k^{a}k_{1}\mid r(t)$, where $k_{1}\mid k,k_{1}\nmid m, a\in\mathbb{N}$, then $k^{a-1}k_{1}\mid r(t+1),\ldots, k_{1}\mid r(t+a),k\nmid r(t+a),k_{1}\nmid r(t+a+1)$.
\end{example}

Example \ref{Ex1} shows the necessity of factorizing dimensions of system matrix $A$ and initial space $\mathcal{V}_{p}$ before computing state dimensions of system (\ref{Eq3}).
Assume $k=k_{1}^{\mu_{1}}k_{2}^{\mu_{2}}\cdots k_{\varpi}^{\mu_{\varpi}},
m=m_{1}^{\nu_{1}}m_{2}^{\nu_{2}}\cdots m_{\omega}^{\nu_{\omega}}$, where $k_{i},m_{j},i=1,2,\ldots,\varpi,j=1,2,\ldots,\omega$ are prime numbers.
$p$ is factorized according to the following steps.

(1) Write $p$ in form of $p=k^{\alpha}p'$, where $k\nmid p'$.

(2) Write $p'$ in form of $p'=k_{1}^{\beta_{1}}k_{2}^{\beta_{2}}\cdots k_{\varpi}^{\beta_{\varpi}}p''$, where $k_{i}\nmid p'', i=1,\ldots,\varpi$.
Due to $k\nmid p'$, there is at least one positive integer $l$ such that $\beta_{l}<\mu_{l}$ holds.

(3) Write $p''$ in form of $p''=m_{1}^{\theta_{1}}m_{2}^{\theta_{2}}\cdots m_{\omega}^{\theta_{\omega}}p_{1}$, where $m_{i}\nmid p_{1},i=1,\ldots,\omega$.
If $k_{i}=m_{j}$, then $\theta_{j}=0$, i.e., $m_{j}^{\theta_{j}}=1$.
Hence, we assume $(m_{i},k)=1,i=1,\ldots,\omega$.

To give the expression of state diemsnion $r(t)$ better, some assumptions are presented.

\begin{assumption}\label{Assu2}\rm
Assume $k=k_{1}^{\mu_{1}}k_{2}^{\mu_{2}}\cdots k_{\varpi}^{\mu_{\varpi}},
m=m_{1}^{\nu_{1}}m_{2}^{\nu_{2}}\cdots m_{\omega}^{\nu_{\omega}}$, where $k_{i},m_{j},$
$i=1,2,\ldots,\varpi,j=1,2,\ldots,\omega$ are prime numbers.
Write $p$ in form of
$$p=k^{\alpha}k_{1}^{\beta_{1}}k_{2}^{\beta_{2}}\cdots k_{\varpi}^{\beta_{\varpi}}m_{1}^{\theta_{1}}m_{2}^{\theta_{2}}\cdots m_{\omega}^{\theta_{\omega}}p_{1},$$
where

\begin{minipage}[t]{0.8\linewidth}
$0<d\leq\varpi,\ \ \beta_{i}<\mu_{i},\ i\leq d;\\
\beta_{i}=\tau_{i}\mu_{i}+\eta_{i},\ \ \eta_{i}<\mu_{i},\ i>d,\ \tau_{i}\leq\tau_{j},\ i<j;\\
0<l\leq\omega;\ \ \nu_{i}>\theta_{i},\ i\leq l;\ \ \nu_{i}\leq\theta_{i},\ i>l;\\
(p_{1},km)=1,\ \ (m_{i},k)=1,\ i=1,2,\ldots,\omega;\\
\{\mu_{i},\nu_{j},\alpha,\beta_{i},\theta_{j},i=1,2,\ldots,\varpi,j=1,2,\ldots,\omega\}\subset\mathbb{N}.\\
$
\end{minipage}

\end{assumption}

\begin{theorem}\label{Alg1}\rm
Consider dimension-bounded linear system (\ref{Eq3}) with initial space $\mathcal{V}_{p}$.
Under Assumption \ref{Assu2}, state dimension $r(t)$ can be computed directly.

(1) If $t\leq\alpha$, then state dimension $r(t)$ is
\[r(t)=mk^{\alpha-t}\prod\limits_{i=1}^{\varpi}k_{i}^{\beta_{i}}\prod\limits_{j=l}^{\omega}m_{j}^{\theta_{j}-\nu_{j}}p_{1}.\]

(2) Let $\tau_{d}=0$.
If $\alpha+\tau_{k}<t\leq\alpha+\tau_{k+1},k=d,d+1,\ldots,\varpi-1$, then state dimension $r(t)$ is
\[r(t)=m\prod\limits_{i=k+1}^{\varpi}k_{i}^{(\tau_{i}+\alpha-t)\mu_{i}+\eta_{i}}\prod\limits_{j=l}^{\omega}m_{j}^{\theta_{j}-\nu_{j}}p_{1}.\]

(3) If $t\geq\alpha+\tau_{\varpi}+1$, then state dimension $r(t)$ is
\[r(t)=m\prod\limits_{j=l}^{\omega}m_{j}^{\theta_{j}-\nu_{j}}p_{1}.\]
\end{theorem}

\begin{proof}
These results can be derived by direct computations.
\end{proof}

\begin{remark}\label{Re3}\rm
It is worth noting that the number of multiplication operations of state dimension $r(t)$ is not more than $\alpha+\varpi\beta+\omega\gamma+3$, where $\beta=\max\{\beta_{1},\ldots,$
$\beta_{\varpi}\}$ and $\gamma=\max\{\theta_{1}-\nu_{1},\ldots,\theta_{\omega}-\nu_{\omega}\}$.
\end{remark}

Consider system (\ref{Eq3}) with $A\in\mathcal{M}_{2\times 6}$.
when $x(0)\in\mathcal{V}_{5}$, the invariant time point of system (\ref{Eq3}) is $t^{\ast}=1$.
The invariant time point of system (\ref{Eq3}) is $t^{\ast}=2$ if $x(0)\in\mathcal{V}_{18}$.
It shows that the invariant time point $t^{\ast}$ of system (\ref{Eq3}) depends on the dimension of initial value.
Hence, we denote the invariant time point $t^{\ast}$ as $t^{\ast}(p)$ with $p$ being the dimension of initial value.

\begin{remark}\label{Re1}\rm
From Theorem \ref{Alg1}, the invariant time point $t^{\ast}$ of system (\ref{Eq3}) can be calculated via formula $t^{\ast}(p)=\alpha+\tau_{\varpi}+1$.
\end{remark}

Based on Theorem \ref{Alg1}, a method is given to judge whether $r$ is a reachable dimension of system (\ref{Eq3}).

\begin{corollary}\label{Cor1}\rm
Consider dimension-bounded linear system (\ref{Eq3}) with initial space $\mathcal{V}_{p}$.
Give a vector space $\mathcal{V}_{r}$.
$r$ is a reachable dimension of system (\ref{Eq3}), if one of the following conditions holds.

(1) $r=r^{\ast}=m\prod\limits_{j=l}^{\omega}m_{j}^{\theta_{j}-\nu_{j}}p_{1}$.

(2) $\frac{r}{r^{\ast}}=k^{\alpha-t}\prod\limits_{i=1}^{\varpi}k_{i}^{\beta_{i}},t\leq\alpha$.

(3) $\frac{r}{r^{\ast}}=\prod\limits_{i=k+1}^{\varpi}k_{i}^{(\tau_{i}+\alpha-t)\mu_{i}+\eta_{i}}, \alpha+\tau_{k}<t\leq\alpha+\tau_{k+1},k=d,d+1,\ldots,\varpi-1$.
\end{corollary}

\subsection{Annihilator polynomial}

For investigating the reachability after the invariant time point $t^{\ast}$, this subsection discusses annihilator polynomials.
To begin with, the definition of annihilator polynomial is provided.

\begin{definition}\label{Def5}\rm\cite{Chengdaizhan2019b}
Give a matrix $A\in\mathcal{M}$, a vector $x\in\mathcal{V}$ and a polynomial
\begin{equation}\label{Eq6}
q(z)=z^{n}+c_{n-1}z^{n-1}+\cdots+c_{1}z+c_{0}.
\end{equation}
(1) $q(z)$ is called an $A$-annihilator of $x$, if
\small
\[
q(A)\vec{\ltimes}x:=A^{n}\vec{\ltimes}x\vec{\rotatebox[origin=c]{90}{$\mp$}}c_{n-1}A^{n-1}\vec{\ltimes}x\vec{\rotatebox[origin=c]{90}{$\mp$}}\cdots\vec{\rotatebox[origin=c]{90}{$\mp$}}c_{1}A\vec{\ltimes}x\vec{\rotatebox[origin=c]{90}{$\mp$}}c_{0}x=\mathbf{0}.
\]
(2) If $q(z)$ is the $A$-annihilator of $x$ with minimum degree, then $q(z)$ is called the minimum $A$-annihilator of $x$.
\end{definition}

\begin{remark}\label{Re5}\rm
Consider polynomial (\ref{Eq6}).
The positive integer $n$ is called the degree of polynomial (\ref{Eq6}).
For a given matrix $A\in\mathcal{M}$ and a vector $x\in\mathcal{V}$,
if $q(z)$ is the $A$-annihilator of $x$ with minimum degree, then $q(z)$ is an $A$-annihilator of $x$, and each polynomial with degree being less than $n$ is not an $A$-annihilator of $x$.
\end{remark}

Give a matrix $A\in\mathcal{M}_{m\times km}$.
For each $x\in\mathcal{V}$,
Corollary 256 of \cite{Chengdaizhan2019b} pointed out that there exists at least one $A$-annihilator of $x$.
Based on Example 260 of \cite{Chengdaizhan2019b}, we present a constructive proof of this result, which is shown in the following proposition.


\begin{proposition}\label{Pro2}\rm
Give a matrix $A\in\mathcal{M}_{m\times km}$.
For each vector $x\in\mathcal{V}$, there exists an integer $i\in\mathbb{N}$ and a set of coefficients $c_{0},c_{1},\ldots,c_{i-1}$ such that polynomial $q(z)=z^{i}+c_{i-1}z^{i-1}+\cdots+c_{1}z+c_{0}$ is the minimal $A$-annihilator of $x$.
\end{proposition}

\begin{proof}
If $x=\mathbf{0}$, then the minimal $A$-annihilator of $x$ is $q(z)=1$.
Otherwise, the minimal $A$-annihilator of $x$ is obtained by the following steps.\\
\indent
Let $x_{0}=x\in\mathcal{V}_{r_{0}}$ and $x_{1}=A\vec{\ltimes}x_{0}\in\mathcal{V}_{r_{1}}$.
Then we calculate $y_{0}=x_{0}\otimes\mathbf{1}_{t_{1}/r_{0}}$ and $y_{1}=x_{1}\otimes\mathbf{1}_{t_{1}/r_{1}}$, where $t_{1}=\mathrm{lcm}(r_{0},r_{1})$.
If there exist $c'_{0},c'_{1}\in\mathbb{R}$ and $c'_{1}\neq 0$ such that $c'_{1}A\vec{\ltimes}x_{0}\vec{\rotatebox[origin=c]{90}{$\mp$}}c'_{0}x_{0}=c'_{1}x_{1}\vec{\rotatebox[origin=c]{90}{$\mp$}}c'_{0}x_{0}=c'_{1}y_{1}+c'_{0}y_{0}=\mathbf{0}$, then the minimal $A$-annihilator of $x_{0}$ is $q(z)=z+c_{0}$, where $c_{0}=\frac{c'_{0}}{c'_{1}}$.
Otherwise, repeat this process until coefficients $c'_{0},c'_{1},\ldots,c'_{i},c'_{i}\neq 0$ satisfying $c'_{i}A^{i}\vec{\ltimes}x_{0}\vec{\rotatebox[origin=c]{90}{$\mp$}}c'_{i-1}A^{i-1}\vec{\ltimes}x_{0}\vec{\rotatebox[origin=c]{90}{$\mp$}}\cdots\vec{\rotatebox[origin=c]{90}{$\mp$}}c'_{1}A\vec{\ltimes}x_{0}\vec{\rotatebox[origin=c]{90}{$\mp$}}c'_{0}x_{0}=\mathbf{0}$
are derived.
Since \cite{Chengdaizhan2019b} proved that $x_{0},A\vec{\ltimes}x_{0},\ldots,A^{i}\vec{\ltimes}x_{0}$ enter an $A$-invariant space at finite steps, $\{c'_{0},c'_{1},\ldots,c'_{i}\}$ is a finite set.
Therefore, the minimal $A$-annihilator of $x_{0}$ is $q(z)=z^{i}+c_{i-1}z^{i-1}+\cdots+c_{1}z+c_{0}$, where $c_{j}=\frac{c'_{j}}{c'_{i}},j=0,1,\ldots,i-1$.
\end{proof}

From the proof of Proposition \ref{Pro2}, for any $x,y\in\mathcal{V}$ satidfying $x\neq y$, the annihilator polynomial of $x$ may not be equal to that of $y$.
Thus, we define the annihilator polynomial of subset $U\subset\mathcal{V}$, which is the generalization of conventional annihilator polynomial.

\begin{definition}\label{Def6}\rm
Give a matrix $A\in\mathcal{M}_{m\times km}$, a subset $U\subset\mathcal{V}$ and a polynomial
\[
q(z)=z^{n}+c_{n-1}z^{n-1}+\cdots+c_{1}z+c_{0}.
\]

(1) $q(z)$ is called an $A$-annihilator of $U$, if
$q(z)$ is an $A$-annihilator of each vector $x\in U$.

(2) If $q(z)$ is the $A$-annihilator of $U$ with minimum degree, then $q(z)$ is called the minimum $A$-annihilator of $U$.
\end{definition}

Give a matrix $A\in\mathcal{M}_{n\times n}$.
If $q(z)$ is an $A$-annihilator of $\mathcal{V}_{n}$,
then $q(z)$ is an $A$-annihilator of $\delta_{n}^{i},\ i=1,2,\ldots,n$, i.e., $q(A)\delta_{n}^{i}=\mathbf{0},\ i=1,2,\ldots,n$.
$q(z)$ is a conventional annihilator polynomial because
$\mathbf{0}=[q(A)\delta_{n}^{1}\ q(A)\delta_{n}^{2}\ \cdots\ q(A)\delta_{n}^{n}]$
$=q(A)I_{n}=q(A).$
It implies that an $A$-annihilator of $\mathcal{V}_{n}$ is the generalization of conventional annihilator polynomial.
The following proposition proposes an approach to derive the minimal annihilator polynomial of $\mathcal{V}_{n}$.

\begin{proposition}\label{Pro3}\rm
Give a matrix $A\in\mathcal{M}_{m\times km}$.
If $q_{i}(z)$ is the minimum $A$-annihilator of $\delta_{n}^{i},\ i=1,2,\ldots,n$,
then $q(z)=\mathrm{lcm}(q_{1}(z),q_{2}(z),\ldots,q_{n}(z))$ is the minimum $A$-annihilator of $\mathcal{V}_{n}$.
\end{proposition}

\begin{proof}
Obviously, $q(z)$ is an $A$-annihilator of $\delta_{n}^{i},i=1,2,\ldots,n$.
For each $x\in\mathcal{V}_{n}$, denote $x=k_{1}\delta_{n}^{1}+k_{2}\delta_{n}^{2}+\cdots+k_{n}\delta_{n}^{n}$.
Then
$q(A)\vec{\ltimes}x=q(A)\vec{\ltimes}(k_{1}\delta_{n}^{1}+k_{2}\delta_{n}^{2}+\cdots+k_{n}\delta_{n}^{n}) =k_{1}q(A)\vec{\ltimes}\delta_{n}^{1}+k_{2}q(A)\vec{\ltimes}\delta_{n}^{2}+\cdots+k_{n}q(A)\vec{\ltimes}\delta_{n}^{n}=\mathbf{0}.$
Hence, $q(z)$ is an $A$-annihilator of $x$.
Due to the arbitrariness of $x$, $q(z)$ is an $A$-annihilator of $\mathcal{V}_{n}$.\\
\indent
Assume $f(z)$ is the minimum $A$-annihilator of $\mathcal{V}_{n}$.
It is obvious that $f(z)\mid q(z)$.
Since $q_{i}(z)$ is the minimum $A$-annihilator of $\delta_{n}^{i},\ i=1,2,\ldots,n$, one sees $q_{i}(z)\mid f(z),\ i=1,2,\ldots,n$.
That is, $f(z)$ is a common multiple of $q_{i}(z),\ i=1,2,\ldots,n$.
Thus, we have $q(z)\mid f(z)$, which means $f(z)=aq(z),a\in\mathbb{R},a\neq 0$.
From the proof above, we find that $q(z)$ is the minimum $A$-annihilator of $\mathcal{V}_{n}$.
\end{proof}

\begin{remark}\label{Re4}\rm
For linear space $U=\mathrm{span}\{x_{1},x_{2},\ldots,x_{n}\}$, if $q_{i}(z)$ is the minimal $A$-annihilator of $x_{i},i=1,2,\ldots,n$, then $q(z)=\mathrm{lcm}(q_{1}(z),\ldots,q_{n}(z))$ is the minimal $A$-annihilator of $U$.
\end{remark}

\subsection{Reachable subsets}

Reachable subsets are computed in this subsection.
Since V-product is linear with respect to the second variable, it is easy to prove that the $t$-step reachable subset is a linear space.

\begin{theorem}\label{Cor2}\rm
Consider dimension-bounded linear system (\ref{Eq3}) with initial space $\mathcal{V}_{p}$.
The $t$-step reachable subsset of the system is a linear space.
Moreover, the $t$-step reachable subset is called the $t$-step reachable subspace and calculated via
$$R_{t}=\mathrm{span}\{A^{t}\vec{\ltimes}\delta_{p}^{1},A^{t}\vec{\ltimes}\delta_{p}^{2},\ldots,A^{t}\vec{\ltimes}\delta_{p}^{p}\}.$$
\end{theorem}

\begin{remark}\label{Re6}\rm
To derive the $t$-step reachable subspace of dimension-bounded linear systems, up to $pt$ times matrix multiplications are required according to Theorem \ref{Cor2}.
For each matrix multiplication, the number of multiplication operations is not more than $\max\{2m^{3}k^{2t-1},\frac{3a^{2}}{k^{t}}\}$, where $a=\mathrm{lcm}(k^{t}m,p)$.
\end{remark}

On the basis of Theorem \ref{Cor2}, a necessary and sufficient condition about $x\in R_{t}$ is provided.

\begin{theorem}\label{Th2}\rm
Consider dimension-bounded linear system (\ref{Eq3}) with initial space $\mathcal{V}_{p}$.
A state $x$ is the $t$-step reachable if and only if
\[
\mathrm{rank}(x,A^{t}\vec{\ltimes}\delta_{p}^{1},A^{t}\vec{\ltimes}\delta_{p}^{2},\ldots,A^{t}\vec{\ltimes}\delta_{p}^{p})=\dim R_{t},
\]
where $\dim R_{t}$ is the dimension of the $t$-step reachable subspace.
\end{theorem}

\begin{remark}\label{Re2}\rm
For a dimension-unbounded linear system, results about the $t$-step reachable subspace also hold.
\end{remark}

Suppose system (\ref{Eq3}) reaches the invariant space $\mathcal{V}_{r^{\ast}}$ at the invariant time point $t^{\ast}$.
Then $\bigcup_{t\geq t^{\ast}}R_{t}\subset\mathcal{V}_{r^{\ast}}$ is the reachable subset of system (\ref{Eq3}) after time $t^{\ast}$.
What we concern is whether $\bigcup_{t\geq t^{\ast}}R_{t}$ equals $\mathcal{V}_{r^{\ast}}$.
Annihilator polynomials are used to answer this question.
To this end, an approach to obtaining the minimum $A$-annihilator of $\bigcup_{t\geq t^{\ast}}R_{t}$ is given.

\begin{theorem}\label{Th5}\rm
Consider dimension-bounded linear system (\ref{Eq3}) with initial space $\mathcal{V}_{p}$.
If $q_{i}(z)$ is the minimum $A$-annihilator of $A^{t^{\ast}}\vec{\ltimes}\delta_{p}^{i},\ i=1,2,\ldots,p$,
then $q(z)=\mathrm{lcm}(q_{1}(z),q_{2}(z),\ldots,q_{p}(z))$ is the minimum $A$-annihilator of $\bigcup_{t\geq t^{\ast}}R_{t}$.
\end{theorem}

\begin{proof}
The proof is obvious according to Proposition \ref{Pro3}.
\end{proof}

According to the definition of annihilator polynomial of a given subset, a sufficient condition of $\bigcup_{t\geq t^{\ast}}R_{t}\neq\mathcal{V}_{r^{\ast}}$ is proposed.

\begin{theorem}\label{Th4}\rm
Consider dimension-bounded linear system (\ref{Eq3}).
Suppose $q(z)$ is the minimum $A$-annihilator of $\bigcup_{t\geq t^{\ast}}R_{t}$.
If $q(z)$ is not the minimum $A$-annihilator of $\mathcal{V}_{r^{\ast}}$, then $\bigcup_{t\geq t^{\ast}}R_{t}\neq\mathcal{V}_{r^{\ast}}$.
\end{theorem}

\begin{proof}
Since $q(z)$ is not the minimum $A$-annihilator of $\mathcal{V}_{r^{\ast}}$, there exists an $x\in\mathcal{V}_{r^{\ast}}$ such that $q(z)$ is not an $A$-annihilator of $x$.
Thus, $x\not\in\bigcup_{t\geq t^{\ast}}R_{t}$ because $q(z)$ is the minimum $A$-annihilator of $\bigcup_{t\geq t^{\ast}}R_{t}$.
Therefore, we conclude $\bigcup_{t\geq t^{\ast}}R_{t}\neq\mathcal{V}_{r^{\ast}}$.
\end{proof}

From Theorem \ref{Th4}, a necessary condition of $x\in\bigcup_{t\geq t^{\ast}}R_{t}$ is presented.

\begin{corollary}\label{Cor3}\rm
Consider dimension-bounded linear system (\ref{Eq3}).
Suppose $q(z)$ is the minimum $A$-annihilator of $\bigcup_{t\geq t^{\ast}}R_{t}$.
If state $x$ is reachable after time $t^{\ast}$, then $q(z)$ is the minimum $A$-annihilator of $x$.
\end{corollary}

Before ending this subsection, an example is employed to show how to use the minimum annihilator polynomial to discuss reachability of dimension-bounded linear systems.
In addition, this example also illustrates that there is not inclusion relation between reachable subspaces at times $t^{\ast}+i$ and $t^{\ast}+j,\ i\neq j$.
Furthermore, the intersection of $R_{t^{\ast}+i}$ and $R_{t^{\ast}+j}$ may not be an empty set.

\begin{example}\label{Ex3}\rm
Consider dimension-bounded linear system (\ref{Eq3}) with initial space $\mathcal{V}_{3}$, where
\[
A=\left[
\begin{array}{cccc}
1&0&1&1\\
0&1&0&1\\
\end{array}
\right]_{.}
\]
From subsection 4.1, one sees that the system reaches the invariant space $\mathcal{V}_{6}$ after the invariant time point $t^{\ast}=1$.

(1) We calculate the minimum $A$-annihilator of $\bigcup_{t\geq 1}R_{t}$.
Firstly, $A^{j}\vec{\ltimes}\delta_{3}^{i}, i=1,2,3,j=1,2,3,4,5$ can be computed.
Secondly, since $\mathrm{rank}(A\vec{\ltimes}\delta_{3}^{2},A^{2}\vec{\ltimes}\delta_{3}^{2},$
$A^{3}\vec{\ltimes}\delta_{3}^{2})=3$ and $\mathrm{rank}(A\vec{\ltimes}\delta_{3}^{i},A^{2}\vec{\ltimes}\delta_{3}^{i},$
$A^{3}\vec{\ltimes}\delta_{3}^{i},A^{4}\vec{\ltimes}\delta_{3}^{i})=4,\ i=1,3$.
one obtains
\[
\begin{array}{l}
A^{5}\vec{\ltimes}\delta_{3}^{1}=2A^{4}\vec{\ltimes}\delta_{3}^{1}+2A^{3}\vec{\ltimes}\delta_{3}^{1}-2A^{2}\vec{\ltimes}\delta_{3}^{1}-A\vec{\ltimes}\delta_{3}^{1},\\
A^{4}\vec{\ltimes}\delta_{3}^{2}=A^{3}\vec{\ltimes}\delta_{3}^{2}+3A^{2}\vec{\ltimes}\delta_{3}^{2}+A\vec{\ltimes}\delta_{3}^{2},\\
A^{5}\vec{\ltimes}\delta_{3}^{3}=2A^{4}\vec{\ltimes}\delta_{3}^{3}+2A^{3}\vec{\ltimes}\delta_{3}^{3}-2A^{2}\vec{\ltimes}\delta_{3}^{3}-A\vec{\ltimes}\delta_{3}^{3}.\\
\end{array}
\]
Denote the minimum $A$-annihilator of $A\vec{\ltimes}\delta_{3}^{i}$ by $q_{i}(z),\ i=1,2,3$.
It is easy to see that
\[
\begin{array}{l}
q_{1}(z)=z^{4}-2z^{3}-2z^{2}+2z+1,\\
q_{2}(z)=z^{3}-z^{2}-3z-1,\\
q_{3}(z)=z^{4}-2z^{3}-2z^{2}+2z+1.\\
\end{array}
\]
According to Theorem \ref{Th5}, the minimum $A$-annihilator of $\bigcup_{t\geq 1}R_{t}$ is $$q(z)=\mathrm{lcm}(q_{1}(z),q_{2}(z),q_{3}(z))=z^{4}-2z^{3}-2z^{2}+2z+1.$$

(2) From Proposition \ref{Pro3}, we derive the minimum $A$-annihilator of $\mathcal{V}_{6}$ is $f(z)=z^{6}-2z^{5}-2z^{4}+2z^{3}+z^{2}$.
Because $f(z)=z^{2}q(z)$, we have $\bigcup_{t\geq 1}R_{t}\neq\mathcal{V}_{6}$.

(3) 
Take a subset $U=\mathrm{span}\{A\vec{\ltimes}\delta_{3}^{1},A\vec{\ltimes}\delta_{3}^{2}\}$.
The minimum $A$-annihilator of $U$ is $q(z)$.
Besides, $q(z)$ is also the minimum $A$-annihilator of $A\vec{\ltimes}\delta_{3}^{3}$.
Since $\mathrm{rank}(A\vec{\ltimes}\delta_{3}^{1},$
$A\vec{\ltimes}\delta_{3}^{2},A\vec{\ltimes}\delta_{3}^{3})=3>\dim U,$
one knows $A\vec{\ltimes}\delta_{3}^{3}\not\in U$, which implies that the condition that $q(z)$ is the minimum $A$-annihilator of $A\vec{\ltimes}\delta_{3}^{3}$ is not a sufficient condition of $A\vec{\ltimes}\delta_{3}^{3}\in U$.
Furthermore, for any $x\in\mathcal{V}_{6}$,
the condition that $q(z)$ is the minimum $A$-annihilator of $x$ is a necessary but not sufficient condition of $x\in\bigcup_{t\geq 1}R_{t}$.

(4) Take $y_{1}=[2\ 2\ 3\ 2\ 1\ 1]^{T}$.
$y_{1}=A^{2}\vec{\ltimes}\delta_{3}^{2}$ implies $y_{1}\in R_{2}$.
Besides, one has $y_{1}=A\vec{\ltimes}\delta_{3}^{1}+A\vec{\ltimes}\delta_{3}^{3}$, which means $y_{1}\in R_{1}$.
Thus, we conclude $R_{1}\cap R_{2}\neq\emptyset$.
Take $y_{2}=[3\ 3\ 3\ 2\ 3\ 3]^{T}$ and $y_{3}=[0\ 0\ 1\ 1\ -1\ -1]^{T}$.
Because $y_{2}=A^{2}\vec{\ltimes}\delta_{3}^{3}$ and $y_{3}=A\vec{\ltimes}\delta_{3}^{1}-A\vec{\ltimes}\delta_{3}^{2}$,
one sees $y_{2}\in R_{2}$ and $y_{3}\in R_{1}$.
In addition, $y_{2}\not\in R_{1}$ and $y_{3}\not\in R_{2}$ hold because of $\mathrm{rank}(y_{2},A\vec{\ltimes}\delta_{3}^{1},A\vec{\ltimes}\delta_{3}^{2},A\vec{\ltimes}\delta_{3}^{3})=4$ and $\mathrm{rank}(y_{3},A^{2}\vec{\ltimes}\delta_{3}^{1},A^{2}\vec{\ltimes}\delta_{3}^{2},A^{2}\vec{\ltimes}\delta_{3}^{3})=4$.
Based on the analysis above, we conclude $R_{1}\not\subset R_{2}$ and $R_{2}\not\subset R_{1}$.
\end{example}

\section{Conclusion}

The reachability of dimension-bounded linear systems has been studied in this paper.
Based on the expression of state dimension at each time,
a method for judging the reachability of a given vector space $\mathcal{V}_{r}$ has been proposed.
Since the $t$-step reachable subset is a linear space, the $t$-step reachable subset has been calculated via the span of vectors $A^{t}\vec{\ltimes}\delta_{p}^{1},\ldots,A^{t}\vec{\ltimes}\delta_{p}^{p}$, where $A$ is a system matrix.
A rank condition has been given to verify the $t$-step reachability of a given state.
For illustrating the relationship between the invariant space and the reachable subset after the invariant time point $t^{\ast}$, annihilator polynomials have been discussed.
The obtained results have shown that $A$-annihilator of vector space $\mathcal{V}_{n}$ is the generalization of conventional annihilator polynomial, where $A$ is a given matrix.
This paper has provided an example to explain the inclusion relation between reachable subsets at times $t^{\ast}+i$ and $t^{\ast}+j$.

In the future work, reachability and controllability of dimension-bounded control systems will be focused on.
Based on the existing work, two problems should be solved.
For a given vector space $\mathcal{V}_{r}$, is there a time $t$ and control inputs such that state dimension $r(t)$ of dimension-bounded control systems is $r$?
For a given state $x\in\mathcal{V}_{r}$, is there a time $t$ and control inputs such that the trajectory of the corresponding dimension-bounded control system can reach $x(t)=\mathbf{0}$ from $x(0)=x$?

\end{document}